\documentclass[aps,pre,reprint]{revtex4-1}

\usepackage{amsmath,amssymb,latexsym,mathtools,algorithm,listings,hyperref}
\hypersetup{breaklinks=true}

\def\[{\begin{equation}}
\def\]{\end{equation}}

\def\dd{d}                                % differential
\def\Z{\mathbb Z}                         % integers
\def\R{\mathbb R}                         % reals

\def\H{\mathcal H}              % Hamiltonian
\def\J{Z}                       % site-site coupling
\def\B{B}                       % external field
\newcommand\set[1]{\mathbf #1}  % collection of N spins
\newcommand\avg[1]{\langle#1\rangle} % ensemble average

\def\dim{d}
\def\twodee{\textsc{2\dim} }
\def\threedee{\textsc{3\dim} }

\newcommand\pd[3][]{
  \ifthenelse{\isempty{#1}}
    {\def\tmp{}}
    {\def\tmp{^#1}}
  \frac{\partial\tmp#2}{\partial#3\tmp}
}

\def\CXX{C\texttt{++}}

\begin{document}

\title{Cluster representations and the Wolff algorithm in arbitrary external fields}
\author{Jaron Kent-Dobias}
\author{James P.~Sethna}
\affiliation{Laboratory of Atomic and Solid State Physics, Cornell University, Ithaca, NY, USA}

\date\today

\begin{abstract}
  We introduce a natural way to extend celebrated spin-cluster Monte Carlo
  algorithms for fast thermal lattice simulations at criticality, like Wolff,
  to systems in arbitrary fields, be they linear magnetic vector fields or
  nonlinear anisotropic ones. By generalizing the `ghost spin' representation
  to one with a `ghost transformation,\!' global invariance to spin symmetry
  transformations is restored at the cost of an extra degree of freedom which
  lives in the space of symmetry transformations. The ordinary
  cluster-building process can then be run on the new representation. We show
  that this extension preserves the scaling of accelerated dynamics in the
  absence of a field for Ising, Potts, and $\mathrm O(n)$ models and
  demonstrate the method's use in modelling the presence of novel nonlinear
  fields. We also provide a \CXX\ library for the method's convenient
  implementation for arbitrary models. 
\end{abstract}

\maketitle

Lattice models are important in the study of statistical physics and phase
transitions. Rarely exactly solvable, they are typically studied by
approximate and numerical methods. Monte Carlo techniques are a common way of
doing this, approximating thermodynamic quantities by sampling the
distribution of system states. These Monte Carlo algorithms are better the
faster they arrive at a statistically independent sample. This becomes a
problem near critical points, where critical slowing down
\cite{wolff_critical_1990} results in power-law divergences of dynamic
timescales.

Celebrated cluster algorithms largely addressed this in the absence of
symmetry-breaking fields by using nonlocal updates \cite{janke_nonlocal_1998}
whose clusters undergo a percolation transition at the critical point of the
system \cite{coniglio_clusters_1980}.  These result in relatively small
dynamic exponents for many spin systems \cite{wolff_comparison_1989,
du_dynamic_2006, liu_dynamic_2014, wang_cluster_1990}, including the Ising,
$\mathrm O(n)$ \cite{wolff_collective_1989}, and Potts
\cite{swendsen_nonuniversal_1987, baillie_comparison_1991} models. These
algorithms rely on the natural invariance of the systems in question under
symmetry transformations on their spins.

Some success has been made in extending these algorithms to systems in certain
external fields by adding a `ghost site' \cite{coniglio_exact_1989} that
returns global rotation invariance to spin Hamiltonians at the cost of an
extra degree of freedom, allowing the method to be used in a subcategory of
interesting fields \cite{alexandrowicz_swendsen-wang_1989, wang_clusters_1989,
ray_metastability_1990}. Static fields have also been applied by including a
separate metropolis or heat bath update step after cluster formation
\cite{destri_swendsen-wang_1992, lauwers_critical_1989,
ala-nissila_numerical_1994}, and other categories of fields have been applied
using replica methods \cite{redner_graphical_1998, chayes_graphical_1998,
machta_replica-exchange_2000}.  Monte Carlo techniques that involve cluster
updates at fixed magnetization have been used to examine quantities at fixed
field by later integrating measured thermodynamic functions
\cite{martin-mayor_cluster_2009, martin-mayor_tethered_2011}.

We show that the scaling of correlation time near the critical point of
several models suggests that the `ghost' approach is a natural one, e.g., that
it extends the celebrated scaling of dynamics in these algorithms at zero
field to various non-symmetric perturbations. We also show, by a redefinition
of the spin--spin coupling in a generic class of spin systems,
\emph{arbitrary} external fields can be treated using cluster methods. Rather
than the introduction of a `ghost spin,\!' our representation relies on
introducing a `ghost transformation,\!' an extra degree of freedom residing on
a `ghost' site coupled to all other sites that takes its values from the
collection of spin symmetry transformations of the base model rather than
resemble the base spins themselves.

We provide an open-source implementation of this method in the form of a \CXX\
library, available at \url{https://git.kent-dobias.com/wolff/}
\cite{kent-dobias_wolff_2018}. Use of this library will be described briefly
within, but extensive documentation is also available at
\url{https://doc.kent-dobias.com/wolff/}.

\section{Clusters Without a Field}

We will pose the problem in a general way, but several specific examples can
be found in Table~\ref{table:models} for concreteness. Let $G=(V,E)$ be a
graph, where the set of vertices $V=\{1,\ldots,N\}$ enumerates the sites of a
lattice and the set of edges $E$ contains pairs of neighboring sites. Let $R$
be a group acting on a set $X$, with the action of group elements $r\in R$ on
elements $s\in X$ denoted $r\cdot s$. $X$ is the set of states accessible by
each spin, and $R$ is the \emph{symmetry group} of $X$. The set $X$ must admit
a measure $\mu$ that is invariant under the action of $R$, e.g., for any
$A\subseteq X$ and $r\in R$, $\mu(r\cdot A)=\mu(A)$.  This trait is shared by
the counting measure on any discrete set, or by any group acting by isometries
on a Riemannian manifold, such as $\mathrm O(n)$ on $S^{n-1}$ in the $\mathrm
O(n)$ models \cite{caracciolo_wolff-type_1993}.  Finally, a subset $R_2$ of
elements in $R$ of order two must act transitively on $X$. This property,
while apparently obscure, is shared by any symmetric space
\cite{loos_symmetric_1969} or by any transitive, finitely generated isometry
group. In fact, all the examples listed here have spin spaces with natural
metrics whose symmetry group is their set of isometries.  We put one spin at
each site of the lattice described by $G$, so that the state of the entire
system is described by elements $\set s\in X\times\cdots\times X=X^N$. 

The Hamiltonian of this system is a function $\H:X^N\to\R$ defined by
\[
  \H(\set s)=-\!\!\!\!\sum_{\{i,j\}\in E}\!\!\!\!\J(s_i,s_j)-\sum_{i\in V}\B(s_i),
\]
where $\J:X\times X\to\R$ couples adjacent spins and $B:X\to\R$ is an external
field. $\J$ must be symmetric in its arguments and invariant under the action
of any element of $R$ applied to the entire lattice, that is, for any $r\in R$
and $s,t\in X$, $\J(r\cdot s,r\cdot t)=\J(s,t)$.  One may also allow $\J$ to
also be a function of edge---for modelling random-bond, long-range, or
anisotropic interactions---or allow $B$ to be a function of site---for
applying arbitrary boundary conditions or modelling random fields. The formal
results of this paper (that the algorithm obeys detailed balance and
ergodicity) hold equally well for these cases, but we will drop the additional
index notation for clarity. Some extensions, like adding strong random fields
or bonds, ultimately prove inefficient \cite{rieger_monte_1995,
redner_graphical_1998}.

\begin{table*}[htpb]
  \begin{tabular}{l||ccccc}
          & Spins ($X$) & Symmetry ($R$)  &  Action ($g\cdot s$)  &
    Coupling ($\J(s,t)$) & Common Field ($B(s)$) \\
    \hline\hline
    Ising         & $\{-1,1\}$    & $\Z/2\Z$                 &  $0\cdot
    s\mapsto s$, $1\cdot s\mapsto -s$ & $st$ & $Hs$ \\
    $\mathrm O(n)$ & $S^{n-1}$ & $\mathrm O(n)$ & $M\cdot s\mapsto Ms$ & $s^{\mathrm T}t$ & $H^{\mathrm T}s$\\
    Potts & $\{1,\ldots,q\}$ & $\mathrm S_n$ & $(i_1,\ldots,i_q)\cdot s=i_s$ & $\delta(s,t)$ & $\sum_mH_m\delta(m,s)$\\
    Clock & $\Z/q\Z$ & $D_n$ & $r_m\cdot s=m+s$, $s_m\cdot
    s=-m-s$ & $\cos(2\pi\frac{s-t}q)$ & $\sum_mH_m\cos(2\pi\frac{s-m}q)$\\
    \textsc{Dgm} & $\Z$ & $D_{\mathrm{inf}}$ & $r_m\cdot s=m+s$, $s_m\cdot s=-m-s$
    & $(s-t)^2$ & $Hs^2$\\
  \end{tabular}
  \caption{Several examples of spin systems and the symmetry groups that act
    on them. Common choices for the spin--spin coupling in these systems and
    their external fields are also given. Other fields are possible, of course:
    for instance, some are interested in modulated fields $H\cos(2\pi k\theta(s))$ for
    integer $k$ and $\theta(s)$ giving the angle of $s$ to some axis applied
    to the $\mathrm O(2)$ model \cite{jose_renormalization_1977}. All models
    listed here have example implementations in the provided \CXX\ library
    \cite{kent-dobias_wolff_2018}.
  }
  \label{table:models}
\end{table*}

Implementation of a model in the provided library is as simple as defining a
class that represents an element of the state space $X$, with default
constructor (and destructor, if necessary), and a class that represents an
element of the group $R$, with default constructor and member functions that
define the action and inverse action of the class on both states and group
elements. Specific details may be found at
\url{https://doc.kent-dobias.com/wolff/models.html}.

The goal of statistical mechanics is to compute expectation values of
observables $A:X^N\to\R$. Assuming the ergodic hypothesis holds (for systems
with broken-symmetry states, it does not), the expected value $\avg A$ of an
observable $A$ is its average over every state $\set s$ in the configuration
space $X^N$ weighted by the Boltzmann probability of that state appearing, or
\[
\avg A
  =\frac{\int_{X^N}A(\set s)e^{-\beta\H(\set s)}\,\dd\mu(\set s)}
  {\int_{X^N}e^{-\beta\H(\set s)}\,\dd\mu(\set s)},
\]
where for $Y_1\times\cdots\times Y_N=Y\subseteq X^N$ the product measure
$\mu(Y)=\mu(Y_1)\cdots\mu(Y_N)$ is the simple extension of the measure on $X$
to a measure on $X^N$. These values are estimated using Monte Carlo techniques
by constructing a finite sequence of states $\{\set{s_1},\ldots,\set{s_M}\}$
such that
\[
  \avg A\simeq\frac1M\sum_{i=1}^MA(\set{s_i}).
\]
Sufficient conditions for this average to converge to $\avg A$ as $M\to\infty$
are that the process that selects $\set{s_{i+1}}$ given the previous states be
Markovian (only depends on $\set{s_i}$), ergodic (any state can be accessed),
and obey detailed balance (the ratio of probabilities that $\set{s'}$ follows
$\set s$ and vice versa is equal to the ratio of weights for $\set s$ and
$\set{s'}$ in the ensemble).

Measurements of observables during Monte Carlo in the provided library are
made by the use of hooks, which are member functions of a measurement class
that are run at designated points during the algorithm's execution and are
provided arbitrary information about the internal state of all relevant objects.
A detailed description of these hooks can be found at
\url{https://doc.kent-dobias.com/wolff/measurement.html}.

While any of several related cluster algorithms can be described for this
system, we will focus on the Wolff algorithm \cite{wolff_collective_1989}. In
the absence of an external field, e.g., $B(s)=0$, the Wolff algorithm proceeds
in the following way.\\

\begin{algorithm}[H]
  \begin{enumerate}
    \item Pick a random site $m_0$ and add it to the stack.

    \item Select a transformation $r\in R_2$ distributed by $f(r\mid m_0,\set s)$.
      Often $f$ is taken as uniform on $R_2$, but it is sufficient for preserving
      detailed balance that $f$ be any function of the seed site $m_0$ and
      $Z(s,r\cdot s)$ for all $s\in\set s$. The flexibility offered by the
      choice of distribution will be useful in situations where the set of spin
      states is infinite.
    \item While the stack isn't empty,
      \begin{enumerate}
        \item pop site $m$ from the stack.
        \item If site $m$ isn't marked, 
          \begin{enumerate}
            \item mark the site.
            \item For every $j$ such that $\{m,j\}\in E$, add site $j$ to the
              stack with probability
              \[
                p_r(s_m,s_j)=\min\{0,1-e^{\beta(\J(r\cdot s_m,s_j)-\J(s_m,s_j))}\}.
                \label{eq:bond_probability}
              \]
            \item Take $s_m\mapsto r\cdot s_m$.
        \end{enumerate}
      \end{enumerate}
  \end{enumerate}
  \caption{Wolff}
  \label{alg:wolff}
\end{algorithm}

When the stack is exhausted, a cluster of connected spins will have been
transformed by the action of $r$. In order for this algorithm to be useful, it
must satisfy ergodicity and detailed balance. Ergodicity is satisfied since we
have ensured that $R_2$ acts transitively on $X$, e.g., for any $s,t\in X$
there exists $r\in R_2$ such that $r\cdot s=t$. Since there is a nonzero
probability that only one spin is transformed and that spin can be transformed
into any state, ergodicity follows. The probability $P(\set s\to\set{s'})$
that the configuration $\set s$ is brought to $\set s'$ by the flipping of a
cluster formed by accepting transformations of spins via bonds $C\subseteq E$
and rejecting transformations via bonds $\partial C\subset E$ is related to
the probability of the reverse process $P(\set{s'}\to\set s)$ by
\begin{widetext}
\[
  \begin{aligned}
    \frac{P(\set s\to\set{s'})}{P(\set{s'}\to\set s)}
    &=\frac{f(r\mid m_0,\set s)}{f(r^{-1}\mid m_0,\set s')}
      \prod_{\{i,j\}\in }\frac{p_r(s_i,s_j)}{p_{r^{-1}}(s_i',s_j')}
      \prod_{\{i,j\}\in\partial C}\frac{1-p_r(s_i,s_j)}{1-p_{r^{-1}}(s'_i,s'_j)}
    =\!\!\!\prod_{\{i,j\}\in\partial C}e^{\beta(\J(s_i',s_j')-\J(s_i,s_j))}
    =\frac{e^{-\beta\H(\set{s'})}}{e^{-\beta\H(\set s)}},
  \end{aligned}
\]
\end{widetext}
whence detailed balance is also satisfied, using $r=r^{-1}$ and $Z(r\cdot
s',s')=Z(r\cdot s,s)$.

The Wolff algorithm is well known to be efficient in sampling many spin models
near and away from criticality, including the Ising, Potts, and $\mathrm O(n)$
models. In general, its efficiency will depend on the system at hand, e.g.,
the structure of the configurations $X$ and group $R$. A detailed discussion
of this dependence for a class of configuration spaces with continuous
symmetry groups can be found in \cite{caracciolo_generalized_1991,
caracciolo_wolff-type_1993}.

This algorithm can be run on a system using the provided library. To construct
a system, you must provide a graph representing the lattice, a temperature,
the spin coupling function $Z$, and the field coupling function $B$. Once
constructed, cluster flips as described in Alg.~\ref{alg:wolff} can be
performed by directly providing seed sites $m_0$ and transformations $r$, or
many in sequence by providing a function that generates random (appropriately
distributed to preserve detailed balance) transformations $r$. The
construction and use of Wolff systems is described at
\url{https://doc.kent-dobias.com/wolff/system.html}.

\section{Adding the field}

This algorithm relies on the fact that the coupling $\J$ depends only on
relative orientation of the spins---global reorientations do not affect the
Hamiltonian. The external field $B$ breaks this symmetry. Fortunately it can be
restored. Define a new graph $\tilde G=(\tilde V,\tilde E)$, where $\tilde
V=\{0,1,\ldots,N\}$ adds the new `ghost' site $0$ which is connected by
\[
  \tilde E=E\cup\big\{\{0,i\}\mid i\in V\big\}
\]
to all other sites.  Instead of assigning the ghost site a spin whose value
comes from $X$, we assign it values in the symmetry group $s_0\in R$, so that
the configuration space of the new model is $R\times X^N$. We introduce the
Hamiltonian $\tilde\H:R\times X^N\to\R$ defined by
\[
\begin{aligned}
  \tilde\H(s_0,\set s)
  &=-\!\!\!\!\sum_{\{i,j\}\in E}\!\!\!\!\J(s_i,s_j)
  -\sum_{i\in V}B(s_0^{-1}\cdot s_i)\\
  &=-\!\!\!\!\sum_{\{i,j\}\in\tilde E}\!\!\!\!\tilde\J(s_i,s_j),
\end{aligned}
\]
where the new coupling $\tilde\J:(R\cup X)\times(R\cup X)\to\R$ is defined for
$s,t\in R\cup X$ by
\[
  \tilde\J(s,t) =
  \begin{cases}
    \J(s,t) & \text{if $s,t\in X$} \\
    B(s^{-1}\cdot t) & \text{if $s\in R$} \\
    B(t^{-1}\cdot s) & \text{if $t\in R$}.
  \end{cases}
  \label{eq:new.z}
\]
The modified coupling is invariant under the action of group elements: for any
$r,s_0\in R$ and $s\in X$,
\[
\begin{aligned}
  \tilde\J(rs_0,r\cdot s)
  &=B((rs_0)^{-1}\cdot (r\cdot s))\\
  &=B(s_0^{-1}\cdot s)
  =\tilde\J(s_0,s)
\end{aligned}
\]
The invariance of $\tilde\J$ to global transformations given other arguments
follows from the invariance properties of $\J$.

We have produced a system incorporating the field function $B$ whose
Hamiltonian is invariant under global rotations, but how does it relate to our
old system, whose properties we actually want to measure? If $A:X^N\to\R$ is
an observable of the original system, we construct an observable $\tilde
A:R\times X^N\to\R$ of the new system defined by
\[
  \tilde A(s_0,\set s)=A(s_0^{-1}\cdot\set s)
\]
whose expectation value in the new system equals that of the original
observable in the old system. First, note that $\tilde\H(1,\set s)=\H(\set
s)$. Since the Hamiltonian is invariant under global rotations, it follows
that for any $g\in R$, $\tilde\H(g,g\cdot\set s)=\H(\set s)$.  Using the
invariance properties of the measure on $X$ and introducing a measure $\rho$
on $R$, it follows that
\[
\begin{aligned}
  \avg{\tilde A}
  &=\frac{
    \int_R\int_{X^N}\tilde A(s_0,\set
    s)e^{-\beta\tilde\H(s_0,\set s)}\,\dd\mu(\set s)\,\dd\rho(s_0)
  } {
    \int_R\int_{X^N}e^{-\beta\tilde\H(s_0,\set s)}\,\dd\mu(\set s)\,\dd\rho(s_0)
  }\\
  &=\frac{
    \int_R\int_{X^N}A(s_0^{-1}\cdot\set
    s)e^{-\beta\tilde\H(s_0,\set s)}\,\dd\mu(\set s)\,\dd\rho(s_0)
  } {
    \int_R\int_{X^N}e^{-\beta\tilde\H(s_0,\set s)}\,\dd\mu(\set s)\,\dd\rho(s_0)
  }\\
  &=\frac{
    \int_R\int_{X^N}A(\set{s'})e^{-\beta\tilde\H(s_0,s_0\cdot\set{s'})}\dd\mu(s_0\cdot\set{s'})\,\dd\rho(s_0)
  } {
    \int_R\int_{X^N}e^{-\beta\tilde\H(s_0,s_0\cdot\set{s'})}\dd\mu(s_0\cdot\set{s'})\,\dd\rho(s_0)
  }\\
  &=\frac{
  \int_R\dd\rho(s_0)}{
  \int_R\dd\rho(s_0)}\frac{\int_{X^N}A(\set{s'})e^{-\beta\H(\set{s'})}\dd\mu(\set{s'})
  }{\int_{X^N}e^{-\beta\H(\set{s'})}\dd\mu(\set{s'})
  }
  =\avg A.
\end{aligned}
\]
Using this equivalence, spin systems in a field may be treated in the
following way.
\begin{enumerate}
  \item Add a site to your lattice adjacent to every other site.
  \item Initialize a `spin' at that site whose value is a representation of a
    member of the symmetry group of your ordinary spins.
  \item Carry out the ordinary Wolff cluster-flip procedure on this new
    lattice, substituting $\tilde\J$ as defined in \eqref{eq:new.z} for $\J$.
\end{enumerate}
Ensemble averages of observables $A$ can then be estimated by sampling the
value of $\tilde A$ on the new system. In contrast with the simpler ghost spin
representation, this form of the Hamiltonian might be considered the `ghost
transformation' representation.

One of the celebrated features of the cluster representation of the Ising and
associated models are the improved estimators of various quantities in the
base model, found by measuring conjugate properties of the clusters themselves
\cite{wolff_lattice_1988}. What of these quantities survive this translation?
As is noted in the formative construction of the cluster representation for
the Ising and Potts models, all estimators involving correlators between spins
are preserved, including correlators with the ghost site
\cite{fortuin_random-cluster_1972}. Where a previous improved estimator
exists, we expect this representation to extend it to finite field, all other
features of the algorithm held constant. For instance, the average cluster
size in the Wolff algorithm is often said to be an estimator for the magnetic
susceptibility in the Ising, Potts, and (with clusters weighted by the
components of their spins along the reflection direction
\cite{hasenbusch_improved_1990}) $\mathrm O(n)$ models, but really what
it estimates is the averaged squared magnetization, which corresponds to the
susceptibility when the average magnetization is zero. At finite field the
latter thing is no longer true, but the correspondence between cluster size
and the squared magnetization continues to hold (see
\eqref{eq:cluster-size-scaling} and Fig.~\ref{fig:cluster_scaling} below).

\section{Examples}
\label{sec:examples}

Several specific examples from Table~\ref{table:models} are described in the
following.

\subsection{The Ising model}

In the Ising model spins are drawn from the set $\{1,-1\}$. Its symmetry group
is $C_2$, the cyclic group on two elements, which can be conveniently
represented by a multiplicative group with elements $\{1,-1\}$, exactly the
same as the spins themselves. The only nontrivial element is of order two, and
is selected every time in the algorithm.  Since the symmetry group and the
spins are described by the same elements, performing the algorithm on the
Ising model in a field is fully described by just using the `ghost spin'
representation.  This algorithm or algorithms based on the same decomposition
of the Hamiltonian have been applied by several researchers
\cite{alexandrowicz_swendsen-wang_1989, wang_clusters_1989,
ray_metastability_1990}. The algorithm has been implemented by one of the
authors in an existing interactive Ising simulator at
\url{https://mattbierbaum.github.io/ising.js}
\cite{bierbaum_ising.js_2016}.

\subsection{The $\mathrm O(n)$ models}
\label{sec:examples:on}

In the $\mathrm O(n)$ model spins are described by vectors on the
$(n-1)$-sphere $S^{n-1}$. Its symmetry group is $\mathrm O(n)$, $n\times n$
orthogonal matrices, which act on the spins by matrix multiplication. The
elements of $\mathrm O(n)$ of order two are reflections about hyperplanes
through the origin and $\pi$ rotations about any axis through the origin.
Since the former generate the entire group, reflections alone suffice to
provide ergodicity. Sampling those reflections uniformly works well at
criticality. The `ghost spin' version of the algorithm has been used to apply
a simple vector field to the $\mathrm O(3)$ model
\cite{dimitrovic_finite-size_1991}. Other fields of interest include
$(n+1)$-dimensional spherical harmonics \cite{jose_renormalization_1977} and
cubic fields \cite{bruce_coupled_1975, blankschtein_fluctuation-induced_1982},
which can be applied with the new method. The method is quickly generalized to
spins whose symmetry groups are other compact Lie groups
\cite{caracciolo_generalized_1991, caracciolo_wolff-type_1993}.

At low temperature or high external vector field selecting reflections
uniformly becomes inefficient because the excitations of the model are spin
waves, in which the magnetization only differs by a small amount between
neighboring spins. Under these conditions, most choices of reflection plane
will cause a change in energy so great that the whole system is always
flipped, resulting in many correlated samples. To ameliorate this, one can
draw reflections from a distribution that depends on how the seed spin is
transformed, taking advantage of the freedom to choose the function $f$ in
Alg.~\ref{alg:wolff}. We implement this in the following way. Say that the state of the
seed of the cluster is $s$. Generate a vector $t$ taken uniformly from the
space of unit vectors orthogonal to $s$. Let the plane of reflection be that
whose normal is $n=s+\zeta t$, where $\zeta$ is drawn from a normal
distribution of mean zero and variance $\sigma$. It follows that the tangent
of the angle between $s$ and the plane of reflection is also distributed
normally with zero mean and variance $\sigma$. Since the distribution of
reflection planes only depends on the angle between $s$ and the plane, and
since that angle is invariant under the reflection, this choice preserves
detailed balance.

\begin{figure}
  \include{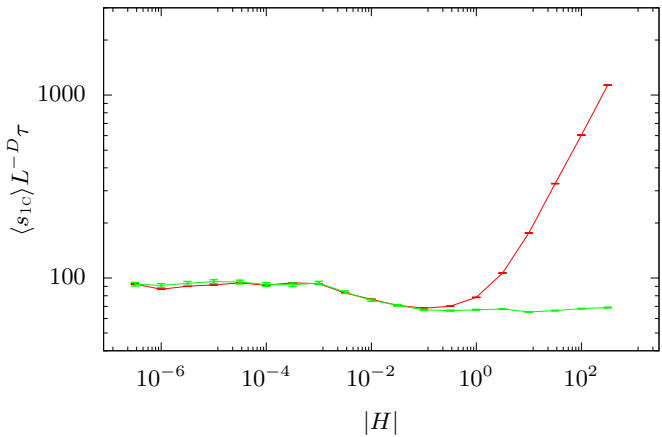}
  \caption{
    The scaled autocorrelation time of the energy $\H$ for the Wolff algorithm
    on a $32\times32\times32$ \textsc{xy} model at its critical temperature as a
    function of applied vector field magnitude $|H|$. Red points correspond to
    reflections sampled uniformly, while the green points represent
    reflections sampled as described in section \ref{sec:examples:on}.
  }
  \label{fig:generator_times}
\end{figure}

The choice of $\sigma$ can be inspired by mean field theory. At high field or
low temperature, spins are likely to both align with the field and each other
and the model is asymptotically equal to a simple Gaussian one, in which in the
limit of large $L$ the expected square angle between neighbors is
\[
  \avg{\theta^2}\simeq\frac{(n-1)T}{D+H/2}.
\]
We take $\sigma=\sqrt{\avg{\theta^2}}/2$. Fig.~\ref{fig:generator_times} shows
the effect of making such a choice on autocorrelation times for the energy for
a critical \threedee \textsc{xy} ($\mathrm O(2)$) model. At small fields both
methods perform the same as zero field Wolff.  Intermediate field values see
efficiency gains for both methods. At large field the uniform sampling method
sees correlation times grow rapidly without bound, while for the sampling
method described here the correlation time crosses over to a constant. A
similar behavior holds for the critical $\mathrm O(3)$ model, though in that
case the constant value the correlation time approaches at large field is
larger than its minimum value (see Fig.~\ref{fig:correlation_time-collapse}).
This behavior isn't particularly worrisome, since the very large field regime
corresponds to correlation lengths comparable to the lattice spacing and is
efficiently simulated by other algorithms.  More detailed discussion on correlation
times and these numeric experiments can be found in section
\ref{sec:performance}.

\subsection{The Potts model}

In the $q$-state Potts model spins are described by elements of
$\{1,\ldots,q\}$. Its symmetry group is the symmetric group $\mathrm S_n$ of
permutations of its elements. The element $(i_1,\ldots,i_q)$ takes the spin
$s$ to $i_s$. There are potentially many elements of order two, but the
two-element swaps alone are sufficient to both generate the group and act
transitively on $\{1,\ldots,q\}$, providing ergodicity.

\subsection{Clock models}

In the $q$-state clock model spins are described by elements of $\Z/q\Z$, the
set of integers modulo $q$. Its symmetry group is the dihedral group
$D_q=\{r_0,\ldots,r_{q-1},s_0,\ldots,s_{q-1}\}$, the group of symmetries of a
regular $q$-gon. The element $r_n$ represents a rotation by $2\pi n/q$, and
the element $s_n$ represents a reflection composed with the rotation $r_n$.
The group acts on spins by permutation: $r_n\cdot m={n+m}\pmod q$ and
$s_n\cdot m={-(n+m)}\pmod q$. This is the natural action of the group on the
vertices of a regular polygon that have been numbered $0$ through $q-1$. The
elements of $D_q$ of order 2 are all reflections and $r_{q/2}$ if $q$ is even,
though the former can generate the latter. While reflections do not
necessarily generate the entire group, their action on $\Z/q\Z$ is transitive
and therefore the algorithm is ergodic.

\subsection{Roughening models}

Though not often thought of as a spin model, roughening of surfaces can be
described in this framework. Spins are described by integers $\Z$ and their
symmetry group is the infinite dihedral group $D_\infty=\{r_i,s_i\mid
i\in\Z\}$, whose action on the spin $j\in\Z$ is given by $r_i\cdot j=i+j$ and
$s_i\cdot j=-i-j$. The elements of order two are reflections $s_i$, whose
action on $\Z$ is transitive. The coupling can be any function of the absolute
difference $|i-j|$.  Because uniform choice of reflection will almost always
result in energy changes so large that the whole system is flipped, it is
better to select random reflections about integers or half-integers close to
the state of the system.  A variant of the algorithm has been applied without
a field whose success relies both on this and another technique
\cite{evertz_stochastic_1991}. They note that detailed balance is still
satisfied if the bond probabilities \eqref{eq:bond_probability} are modified by
adding a constant $0<x\leq1$ with
\[
  p_r(s_m,s_j\mid x)=\min\{0,1-xe^{\beta(\J(r\cdot s_m,s_j)-\J(s_m,s_j))}\}.
\]
When $x<1$ transformations that do not change the energy of a bond can still
activate it in the cluster, which allows nontrival clusters to be seeded when
the height of the starting site is also the plane of reflection. This
modification is likely useful in general for systems with large yet discrete
state spaces.

\section{Performance}
\label{sec:performance}

No algorithm is worthwhile if it doesn't run efficiently. This algorithm,
being an extension of the Wolff algorithm into a new domain, should be
considered successful if it likewise extends the efficiency of the Wolff
algorithm into that domain. Some systems are not efficient under Wolff, and we
don't expect them to fare better when extended in a field. For instance, Ising
models with random fields or bonds technically can be treated with Wolff
\cite{dotsenko_cluster_1991}, but it is not efficient because the clusters
formed do not scale naturally with the correlation length \cite{rieger_monte_1995,
redner_graphical_1998}. Other approaches, like replica methods, should be
relied on instead \cite{redner_graphical_1998, chayes_graphical_1998,
machta_replica-exchange_2000}. 

At a critical point, correlation time $\tau$ scales with system size
$L=N^{-D}$ as $\tau\sim L^z$. Cluster algorithms are celebrated for their
small dynamic exponents $z$. In the vicinity of an ordinary critical point,
the renormalization group predicts scaling behavior for the correlation time
as a function of temperature $t$ and field $h$ of the form
\[
  \tau=h^{-z\nu/\beta\delta}\mathcal T(ht^{-\beta\delta},hL^{\beta\delta/\nu}).
\]
If a given dynamics for a system at zero field results in scaling like $L^z$,
one should expect its natural extension in the presence of a field to scale
roughly like $h^{-z\nu/\beta\delta}$ and collapse appropriately as a function
of $hL^{\beta\delta/\nu}$.

\begin{figure*}
  \include{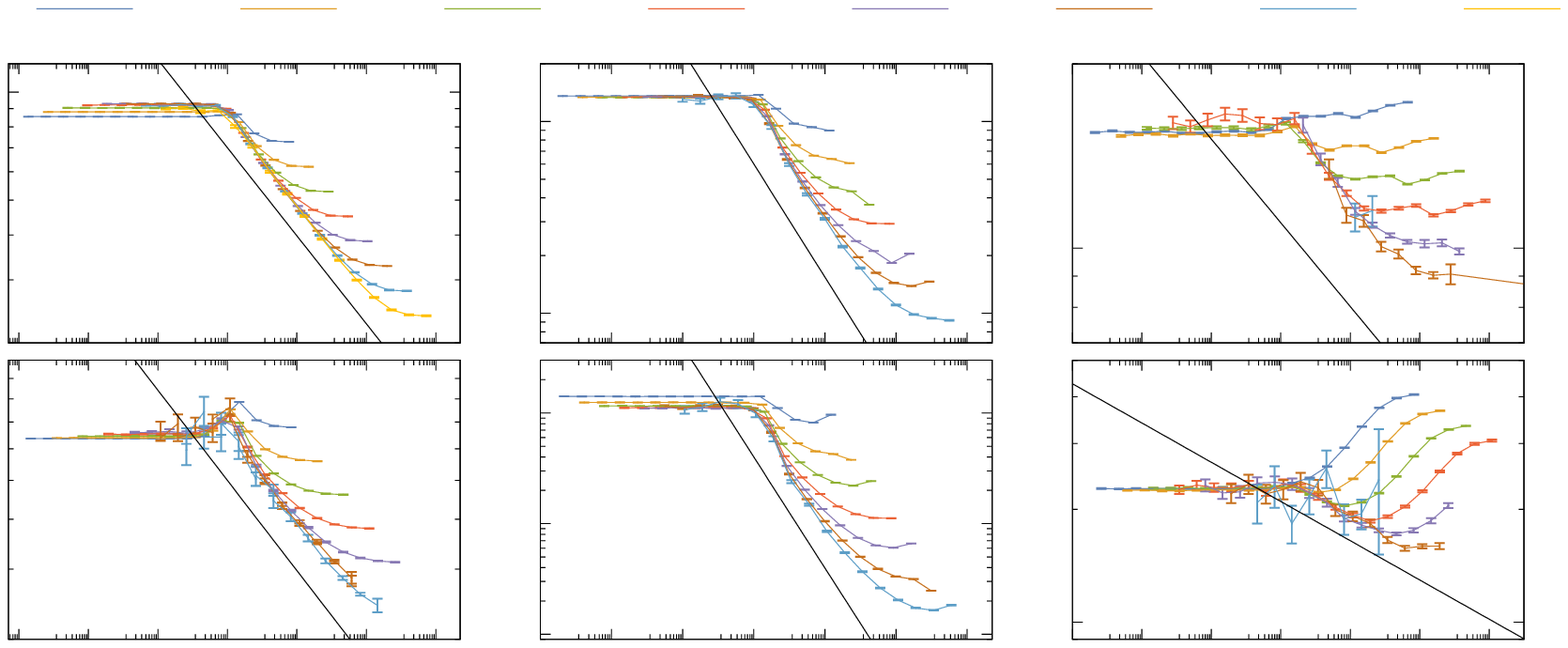}
  \caption{
    Scaling collapse of autocorrelation times $\tau$ for the energy $\H$
    scaled by the average cluster size as a function of external field for
    various models of Table~\ref{table:models}. Critical exponents are
    model-dependent. Colored lines and points depict values as measured by the
    extended algorithm. Solid black lines show a plot proportional to
    $h^{-z\nu/\beta\delta}$ for each model. The dynamic exponents $z$ are
    roughly measured as \twodee Ising: 0.23(5), \threedee Ising: 0.28(5),
    \twodee 3-State Potts: 0.55(5), \twodee 4-State Potts: 0.94(5),
    \threedee O(2): 0.17(5), \threedee O(3): 0.13(5). $\mathrm O(n)$ models
    use the distribution of transformations described in Section
    \ref{sec:examples:on}. The curves stop collapsing at high fields when the
    correlation length falls to near the lattice spacing; here non-cluster
    algorithms can be efficiency used.
  }
  \label{fig:correlation_time-collapse}
\end{figure*}

We measured the autocorrelation time $\tau$ of the energy $\H$ for a variety
of models at critical temperature with many system sizes and canonical fields
(see Table~\ref{table:models} with $h=\beta H$) using standard methods for
obtaining the value and uncertainty from timeseries
\cite{ossola_dynamic_2004}. Since the computational effort expended in each
step of the algorithm depends linearly on the size of the associated cluster,
these values are then scaled by the average cluster size per site
$\avg{s_{\text{\sc 1c}}}/L^D$ to produce something proportional to machine
time per site. The resulting scaling behavior, plotted in
Fig.~\ref{fig:correlation_time-collapse}, is indeed consistent with an
extension to finite field of the behavior at zero field, with an eventual
finite-size crossover to constant autocorrelation time at large field. This
crossover isn't always kind to the efficiency, e.g., in the $\mathrm O(3)$
model, but in the large-field regime where the crossover happens the
correlation length is on the scale of the lattice spacing and better
algorithms exist, like Bortz--Kalos--Lebowitz for the Ising model
\cite{bortz_new_1975}. Also plotted are lines proportional to
$h^{-z\nu/\beta\delta}$, which match the behavior of the correlation times in
the intermediate scaling region as expected. Values of the critical exponents
for the models were taken from the literature \cite{wu_potts_1982,
el-showk_solving_2014, guida_critical_1998} with the exception of $z$ for the
energy in the Wolff algorithm, which was determined for each model by making a
power law fit to the constant low field behavior.  These exponents are
imprecise and are provided in the figure with only qualitative uncertainty.

Since the formation and flipping of clusters is the hallmark of Wolff
dynamics, another way to ensure that the dynamics with field scale like those
without is to analyze the distribution of cluster sizes. The success of the
algorithm at zero field is related to the fact that the clusters formed
undergo a percolation transition at models' critical point.  According to the
scaling theory of percolation \cite{stauffer_scaling_1979}, the distribution
of cluster sizes in a full Swendsen--Wang decomposition---where the whole
system is decomposed into clusters with every bond activated with probability
\eqref{eq:bond_probability}---of the system scales consistently near the
critical point if it has the form
\[
  P_{\text{SW}}(s)=s^{-\tau}f(ts^\sigma,th^{-1/\beta\delta},tL^{1/\nu}).
\]
The distribution of cluster sizes in the Wolff algorithm can be computed from
this using the fact that the algorithm selects clusters with probability
proportional to their size, or
\[
  \begin{aligned}
    \avg{s_{\text{\sc 1c}}}&=\sum_ssP_{\text{\sc
    1c}}(s)=\sum_ss\frac sNP_{\text{SW}}(s)\\
    &=L^{\gamma/\nu}g(ht^{-\beta\delta},hL^{\beta\delta/\nu}).
  \end{aligned}
  \label{eq:cluster-size-scaling}
\]

\begin{figure*}
  \input{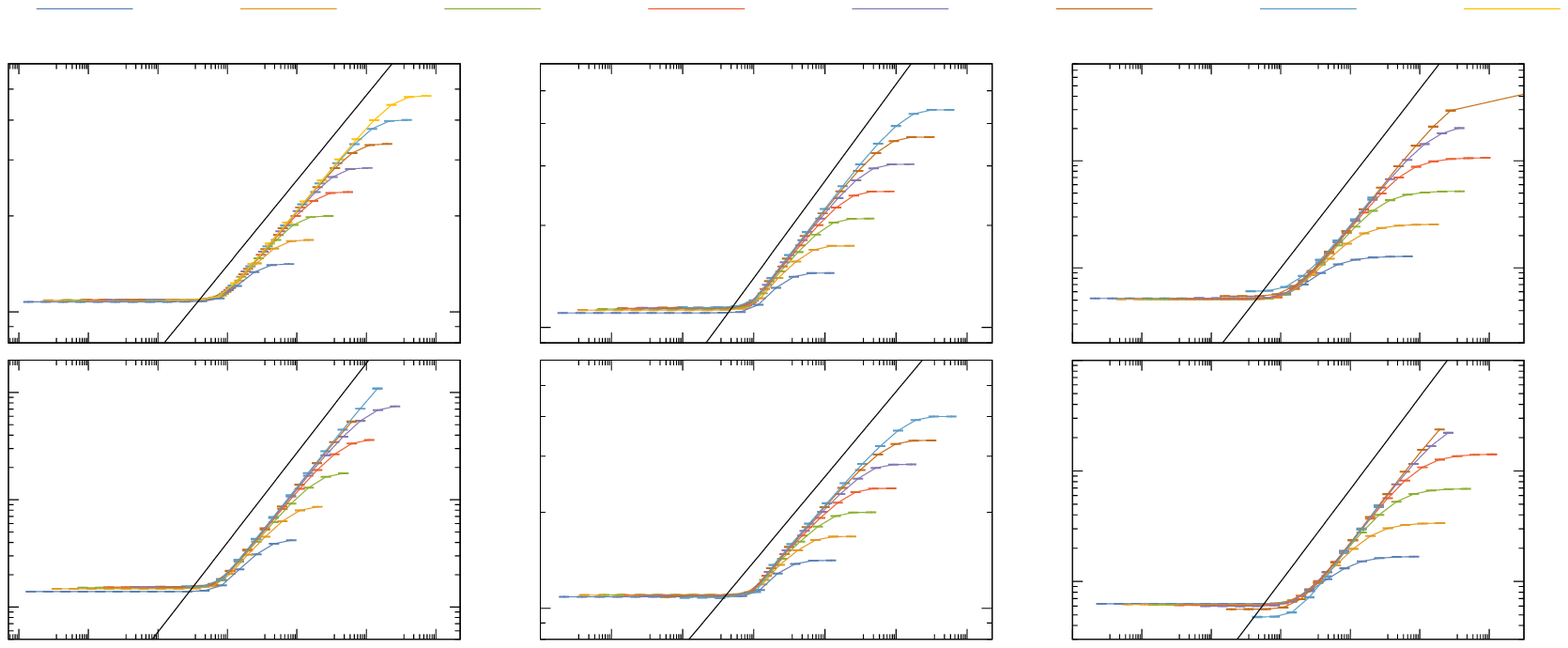}
  \caption{Collapses of rescaled average Wolff cluster size $\avg s_{\text{\sc
    1c}}L^{-\gamma/\nu}$ as a function of field scaling variable
    $hL^{\beta\delta/\nu}$ for a variety of models. Critical exponents
    $\gamma$, $\nu$, $\beta$, and $\delta$ are model-dependant. Colored lines
    and points depict values as measured by the extended algorithm. Solid
    black lines show a plot of $g(0,x)\propto x^{2/\delta}$ for each model.
  }
  \label{fig:cluster_scaling}
\end{figure*}

For the Ising model, an additional scaling relation can be written. Since the
average cluster size is the average squared magnetization, it can be related
to the scaling functions of the magnetization and susceptibility per site by
(with $ht^{-\beta\delta}$ dependence dropped)
\[
  \begin{aligned}
    \avg{s_{\text{\sc 1c}}}
    &=L^{D}\avg{M^2}=\beta\avg\chi+L^{D}\avg{M}^2\\
    &=L^{\gamma/\nu}\big[(hL^{\beta\delta/\nu})^{-\gamma/\beta\delta}\beta \mathcal
      Y(hL^{\beta\delta/\nu},ht^{-\beta\delta})\\
      &\hspace{1em}+(hL^{\beta\delta/\nu})^{2/\delta}\mathcal
    M(hL^{\beta\delta/\nu},ht^{-\beta\delta})\big].
  \end{aligned}
\]
We therefore expect that, for the Ising model, $\avg{s_{\text{\sc
1c}}}L^{-\gamma/\nu}$ should go as $(hL^{\beta\delta/\nu})^{2/\delta}$ for
large argument. We further conjecture that this scaling behavior should hold
for other models whose critical points correspond with the percolation
transition of Wolff clusters.  This behavior is supported by our numeric work
along the critical isotherm for various Ising, Potts, and $\mathrm O(n)$
models, shown in Fig.~\ref{fig:cluster_scaling}. Fields are the canonical ones
referenced in Table~\ref{table:models}. As can be seen, the average
cluster size collapses for each model according to the scaling hypothesis, and
the large-field behavior likewise scales as we expect from the na\"ive Ising
conjecture.

\section{Applying Nonlinear Fields to the xy Model}

Thus far our numeric work has quantified the performance of existing
techniques. Briefly, we demonstrate our general framework in a new way:
harmonic perturbations to the low-temperature {\sc xy}, or \twodee O(2),
model. We consider fields of the form $B_n(s)=h_n\cos(n\theta(s))$, where
$\theta$ is the angle made between $s$ and the $x$-axis. Corrections of these
types are expected to appear in realistic models of systems na\"ively expected
to exhibit Kosterlitz--Thouless critical behavior due to the presence of the
lattice or substrate. Whether these fields are relevant or irrelevant in the
renormalization group sense determines whether those systems spoil or admit
that critical behaviour. Among many fascinating
\cite{jose_renormalization_1977, kankaala_theory_1993,
ala-nissila_numerical_1994, dierker_consequences_1986, selinger_theory_1988}
results that emerge from systems with one or more of these fields applied, it
is predicted that $h_4$ is relevant while $h_6$ is not at some sufficiently
high temperatures below the Kosterlitz--Thouless point
\cite{jose_renormalization_1977}. The sixfold fields are expected to be
present, for instance, in the otherwise Kosterlitz--Thouless-type
two-dimensional melting of argon on a graphite substrate
\cite{zhang_melting_1991}.

\begin{figure}
  \include{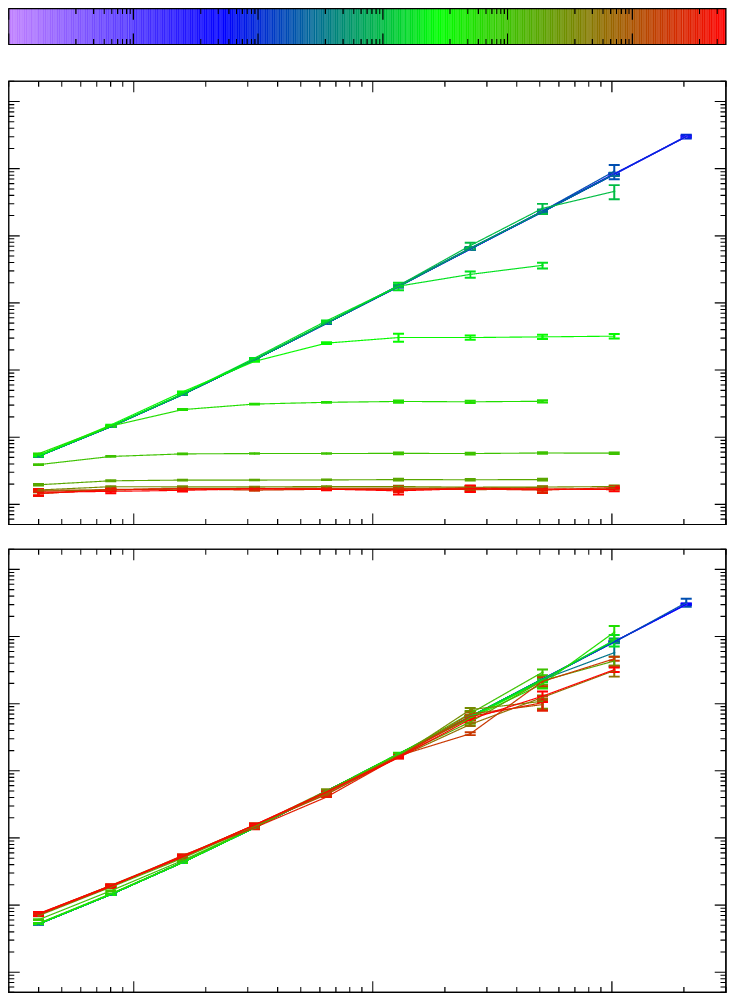}
  \caption{Susceptibilities as a function of system size for a \twodee O(2)
    model at $T=0.7$ and with (top) fourfold symmetric and (bottom) sixfold
    symmetric perturbing fields. Different field strengths are shown in
    different colors.
  }
  \label{fig:harmonic-susceptibilities}
\end{figure}

We made a basic investigation of this result using our algorithm. Since we ran
the algorithm at fairly high fields we did not choose reflections though the
origin uniformly. Instead, we choose the planes of reflection first by
rotating our starting spin by $\pi m/n$ for $m$ uniformly taken from
$1,\ldots,n$ and generating a normal to the plane from that direction as
described in Section \ref{sec:examples:on}. The resulting susceptibilities as
a function of system size are shown for various field strengths in
Fig.~\ref{fig:harmonic-susceptibilities}. In the fourfold case, for each field
strength there is a system size at which the divergence in the susceptibility
is cut off, while for the sixfold case we measured no such cutoff, even up to
strong fields. This conforms to the expected result, that even in a strong field
the sixfold perturbations preserve the critical behavior. Previous work has
used Monte Carlo to investigate similar symmetry-breaking fields and used a hybrid
cluster--metropolis method \cite{ala-nissila_numerical_1994}. To our
knowledge, no application of a direct cluster method has been applied to this
problem before now.

\section{Conclusions}

We have taken several disparate extensions of cluster methods to spin models
in an external field and generalized them to work for any model of a broad
class.  The resulting representation involves the introduction of not a ghost
spin, but a ghost transformation. We provide a \CXX\ library with example
implementations of all models described here \cite{kent-dobias_wolff_2018}. We
provided evidence that algorithmic extensions deriving from this method are
the natural way to extend cluster methods in the presence of a field, in the
sense that they appear to reproduce the scaling of dynamic properties in a
field that would be expected from renormalization group predictions.

In addition to uniting several extensions of cluster methods under a single
description, our approach allows the application of fields not possible under
prior methods. Instead of simply applying a spin-like field, this method
allows for the application of \emph{arbitrary functions} of the spins. For
instance, theoretical predictions for the effect of symmetry-breaking
perturbations on spin models can be tested numerically
\cite{jose_renormalization_1977, blankschtein_fluctuation-induced_1982,
bruce_coupled_1975, manuel_carmona_$n$-component_2000}.

\begin{acknowledgments}
  This work was supported by NSF grant NSF DMR-1719490.
\end{acknowledgments}

\appendix

\section{Example Ising Implementation}

Provided below is an example implementation of the Ising model using the
provided \CXX\ library. The example is also included with the library source,
along with several other more complicated ones \cite{kent-dobias_wolff_2018}.
The routine defines an Ising class that acts as both spin and symmetry group
and a measurement class that provides simple hooks for computing the average
cluster size. The canonical Ising couplings are defined, a square lattice is
initialized, the Wolff system is initialized, and the algorithm is run for a
designated number of cluster flips.

\lstset{language=C++, basicstyle=\footnotesize, frame=single}
\begin{lstlisting}[breaklines]
#include <iostream>
#include <chrono>

#include <wolff.hpp>

using namespace wolff;

class ising_t {
  public:
    int s;

    ising_t() : s(1) {};
    ising_t(int i) : s(i) {};

    ising_t act(const ising_t& x) const {
      return ising_t(s * x.s);
    }

    ising_t act_inverse(const ising_t& x) const {
      return this->act(x);
    }
};

class measure_clusters : public measurement<ising_t, ising_t> {
  private:
    v_t C;

  public:
    double Ctotal;

    measure_clusters() { Ctotal = 0; }

    void pre_cluster(N_t, N_t, const system<ising_t, ising_t>&, v_t, const ising_t&) { C = 0; }

    void plain_site_transformed(const system<ising_t, ising_t>&, v_t, const ising_t&) { C++; }

    void post_cluster(N_t, N_t, const system<ising_t, ising_t>&) { Ctotal += C; }
};

int main(int argc, char *argv[]) {
  // set defaults
  N_t N = (N_t)1e3;
  D_t D = 2;
  L_t L = 128;
  double T = 2.26918531421;
  double H = 0.01;

  // define the spin-spin coupling
  std::function <double(const ising_t&, const ising_t&)> Z =
    [](const ising_t& s1, const ising_t& s2) -> double {
      return (double)(s1.s * s2.s);
    };

  // define the spin-field coupling
  std::function <double(const ising_t&)> B =
    [=](const ising_t& s) -> double {
      return H * s.s;
    };

  // initialize the lattice
  graph G(D, L);

  // initialize the system
  system<ising_t, ising_t> S(G, T, Z, B);

  // define function that generates self-inverse rotations
  std::function <ising_t(std::mt19937&, const system<ising_t, ising_t>&, v_t)> gen_R =
    [] (std::mt19937&, const system<ising_t, ising_t>&, v_t) -> ising_t {
      return ising_t(-1);
    };

  // initailze the measurement object
  measure_clusters A;

  // initialize the random number generator
  auto seed = std::chrono::high_resolution_clock::now().time_since_epoch().count();
  std::mt19937 rng{seed};

  // run wolff N times
  S.run_wolff(N, gen_R, A, rng);

  // print results
  std::cout << "The average cluster size per site was " << (A.Ctotal / N) / S.nv << ".\n";

  // exit
  return 0;
}
\end{lstlisting}

\bibliography{monte-carlo}

\end{document}